\documentclass[aps,prb,showpacs,amsmath,amssymb,twocolumn,superscriptaddress]{revtex4-1}

\usepackage{graphicx}
\usepackage{epsfig}
\usepackage{color}
\usepackage{amsmath}

\begin{document}

\title{Collective amplitude mode fluctuations in a flat band superconductor}

\author{V. J. Kauppila}
\affiliation{O.V. Lounasmaa Laboratory, Aalto University, P.O. Box 15100, FI-00076 AALTO, Finland}
\email[]{ville.kauppila@aalto.fi}
\author{T. Hyart}
\affiliation{University of Jyvaskyla, Department of Physics and Nanoscience Center, P.O. Box 35 (YFL), FI-40014 University of Jyv\"askyl\"a, Finland}
\author{T.~T.~Heikkil\"a}
\affiliation{University of Jyvaskyla, Department of Physics and Nanoscience Center, P.O. Box 35 (YFL), FI-40014 University of Jyv\"askyl\"a, Finland}

\newcommand{\tmpnote}[1]%
   {\begingroup{\it (FIXME: #1)}\endgroup}
   \newcommand{\comment}[1]%
       {\marginpar{\tiny C: #1}}

\date{\today}

\begin{abstract}
We study the fluctuations of the amplitude (i.e. the Higgs-Anderson) mode in a superconducting system of coupled Dirac particles proposed as a model for possible surface or interface superconductivity in rhombohedral graphite. We show that the absence of Fermi energy and vanishing of the excitation gap of the collective amplitude mode in the model leads to a large fluctuation contribution to thermodynamic quantities such as the heat capacity. As a consequence, the mean-field theory becomes inaccurate indicating that the interactions lead to a strongly correlated state. We also present a microscopic derivation of the Ginzburg-Landau theory corresponding to this model.
\end{abstract}


\maketitle

\section{Introduction}

A topological flat band spectrum can emerge as a surface or interface state of topological semimetals \cite{heikkila2011flat}. Such a spectrum has been proposed to occur in rhombohedrally stacked graphite \cite{guinea2006electronic, PhysRevB.83.220503, PhysRevB.87.140503}, at dislocation interfaces in bernal stacked graphite \cite{esquinazi2014superconductivity}, edges of zig-zag graphene \cite{ryu2002topological} and at the interfaces in topological crystalline insulators \cite{tang2014strain}. Such a state has a singular density of states which can, in the presence of an attractive pairing between the particles, lead to superconductivity with an unusually high critical temperature \cite{khodel1990superfluidity,Miyahara07}. This mechanism has been suggested as an explanation to the unusually high critical temperature $T_c$ in some IV-VI semiconductor heterostructures \cite{murase1986superconducting, fogel2006direct, yuzephovich2008interfacial} and in various graphite based materials \cite{scheike2013granular, scheike2012can, ballestar2013josephson}. In the Hubbard model, related work has also been done to study superconductivity in the presence of a flat band \cite{iglovikov2014superconducting} and with Fermi energy close to Van Hove singularities \cite{hirsch1986enhanced}.

\begin{figure}
\centering
\includegraphics[width=.99\columnwidth]{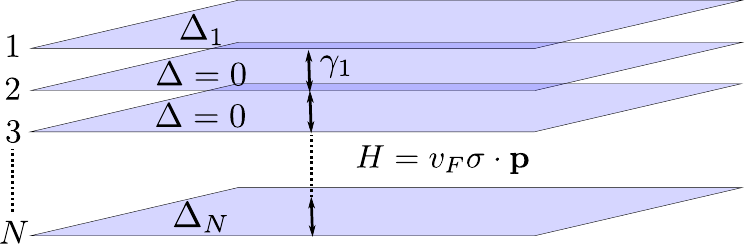}
\caption{\small Schematic figure of the model which consists of $N$ parts with Dirac spectrum $H=v_F \mathbf{\sigma} \cdot \mathbf{p}$ coupled via coupling strength $\gamma_1$. This results in an effective low-energy theory described by the action \eqref{action1}. The superconductivity is localized at the surfaces where the Cooper pair field \cite{footnote1} is  given by $\Delta_{1,N}$.}
\label{fig:schematic}
\end{figure}

\begin{figure}
\centering
\includegraphics[width=.99\columnwidth]{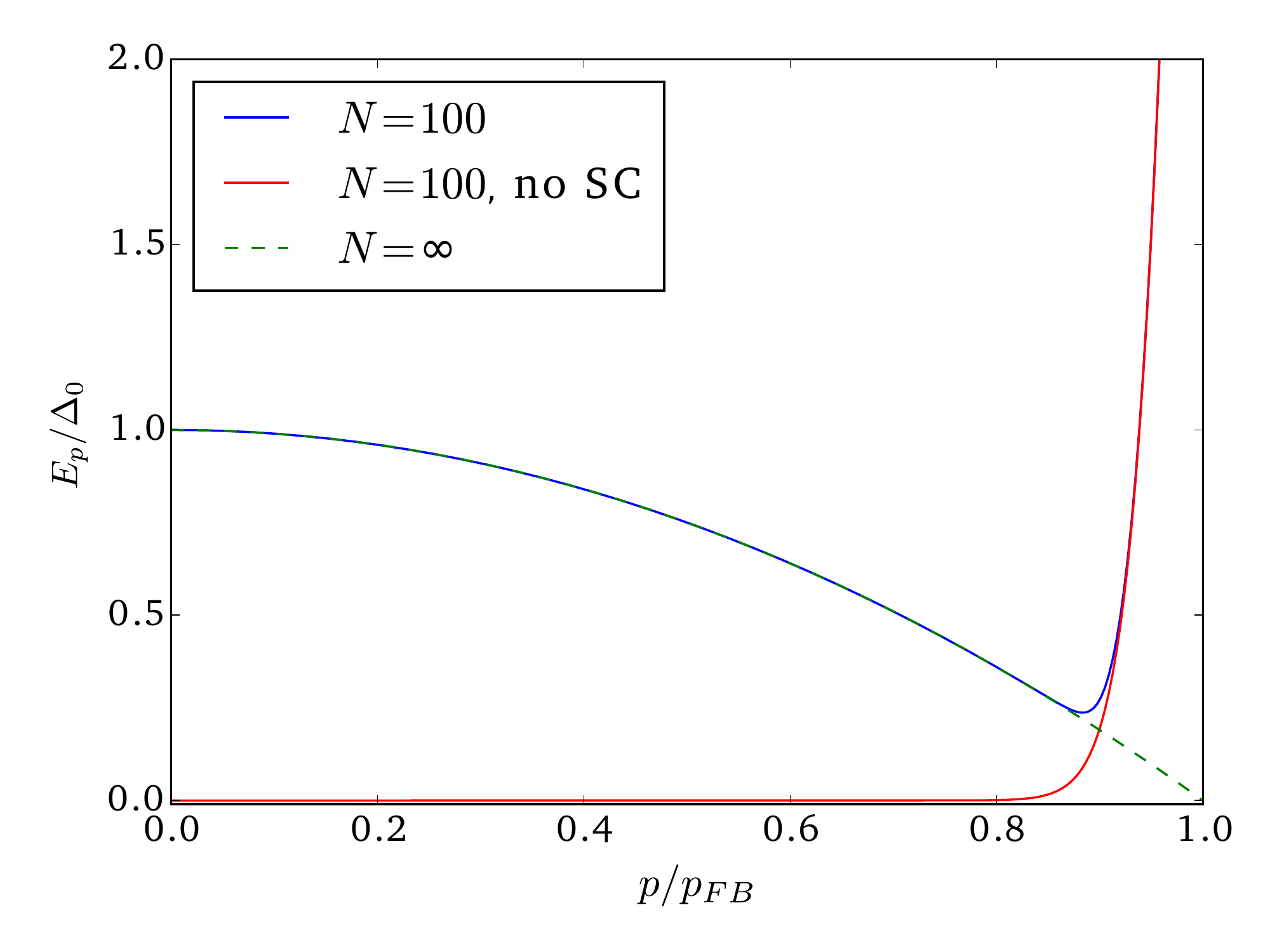}
\caption{\small Fermionic excitation spectrum of the surface state for superconducting (blue line) and normal (black line) state. For finite $N$, the superconducting state has an excitation gap $E_{p,min}$ which closes in the limit $N\rightarrow \infty$.}
\label{fig:spectrum}
\end{figure}

We study here the effect of fluctuations on the superconducting properties of flat band superconductors. We use the particular model proposed in Refs.~\onlinecite{PhysRevB.83.220503} and \onlinecite{PhysRevB.87.140503} and also numerically analyzed in Ref.~\onlinecite{munoz2013tight} for rhombohedrally stacked graphite. The model consists of $N$ individual graphene layers with $s$-wave pairing between the Dirac electrons coupled to form a stack of rhombohedral graphite. The flat band is formed on the surfaces of the stack where the superconductivity also appears (see Fig. \ref{fig:schematic}). A peculiarity of this model is the closing of the gap in the fermionic excitation spectrum at the limit of a large number of layers even with an isotropic $s$-wave mean-field order parameter as can be seen from the surface state spectrum (see Fig. \ref{fig:spectrum})\cite{PhysRevB.87.140503} 
\begin{equation}
E_\mathbf{p}^2 = (1-\mathbf{p}^2 /p_{FB}^2)^2 (\Delta_0^2 + \xi_{\mathbf{p}}^2) ,
\label{spectrum1}
\end{equation}
where $\xi_{\mathbf{p}} = \gamma_1 \vert \mathbf{p} / p_{FB} \vert^N$, $\gamma_1$ is the interlayer coupling constant, $p_{FB} = \gamma_1 / v_F$ is the width of the flat band and $\Delta_0$ is the mean-field order parameter. (We use everywhere $\hbar = k_B = 1$.) This expression is valid for $\vert \mathbf{p} / p_{FB} \vert < 1- 1/N$.\cite{cutoff} Closing of the gap at the flat band edge is due to the fact that there the surface states penetrate into the bulk and surface superconductivity cannot create a gap for the bulk excitations.

Even without making detailed calculations, simple arguments as for why fluctuations are expected to matter in this model can be made. Typically the magnitude of fluctuations is characterized by the Ginzburg number $Gi \sim 1-T^*/T_c$, which is related to the temperature $T^*$ above which the fluctuations dominate the thermodynamical properties. In a 2D BCS superconductor, the Ginzburg number is given by $Gi \sim T_c / E_F$, where $E_F$ is the Fermi energy. For the flat band, $E_F=0$, leaving no other relevant energy scale available to be compared with $T_c$ which suggests\cite{PhysRevB.87.140503} that $Gi \sim 1$. 

We first consider a Ginzburg-Landau theory for this model close to $T_c$. It is of the form expected from symmetry considerations with coupling coefficients that we derive from the microscopic model. Based on this theory, we find that the contribution to the heat capacity from amplitude fluctuations is larger than the mean-field heat capacity jump at the superconducting transition for a wide range of temperatures, resulting in a large Ginzburg number of $Gi = 2/5$ in accord with the dimensional arguments above. Then we consider the same microscopic model, but without making the mean-field approximation. Rather, we employ functional integral calculus to derive the partition function for the system that preserves fluctuations around the mean field up to the Gaussian approximation. We show that the correction from the amplitude mode to the mean-field heat capacity is large even far below the mean-field critical temperature $T_c$. Finally, we numerically show that close to $T=0$ the amplitude mode and the mean field contributions to the heat capacity are approximately equal to each other. This finding can be analytically understood by assuming that at low temperatures the heat capacity is determined by a free boson contribution corresponding to the amplitude mode dispersion and a fermionic contribution due to  Bogoliubov quasiparticles.

\section{Ginzburg-Landau picture}

A simple picture of the fluctuation contribution can be obtained from Ginzburg-Landau theory close to $T_c$. There, a superconductor is described in terms of the Cooper pair field $\Delta$  whose absolute value as well as the gradient are small so that an expansion with respect to them can be made. In the usual approach, only the value of the wave function that minimizes the Ginzburg-Landau free energy is considered, assuming that all the pairs are condensed to this state. The study of fluctuations in superconductors involves considering also Cooper pairs that are on some other states. In the field integral formalism we use below, this translates into taking an integral over different configurations of the Cooper pair fields and weighting different configurations by a factor that is given by the Ginzburg-Landau action when calculating observables.

From symmetry considerations, or from microscopic calculations (see below), we get for a general superconductor consisting of two separate symmetric parts (in our case, the top and the bottom layers of the graphite stack), a Ginzburg-Landau action of the form
\begin{eqnarray}
S_{GL} = \int d^2 x \bigg[ \sum_{i \in\{1, N\}} \big(&& \alpha_1 \vert \Delta_i \vert^2 + \alpha_2 \vert \nabla \Delta_i \vert^2 + \beta \vert \Delta_i \vert^4 \big) \nonumber \\
&& - \gamma (\Delta_1 \bar{\Delta}_N + \Delta_N \bar{\Delta}_1)\bigg] .
\label{glaction}
\end{eqnarray}
Here $\gamma$ is the Josephson coupling coefficient between the two parts characterized by amplitudes $\Delta_1$ and $\Delta_N$. $N$ in the index is the number of layers and it is assumed to be large. The partition function, from which thermodynamic quantities can be calculated, is then given by
\begin{equation}
\mathcal{Z} = \int \Big(\prod_{i \in \{1, N \}} \mathcal{D} \Delta_i \mathcal{D} \bar{\Delta}_i \Big) e^{-S_{GL}(\Delta_i, \bar{\Delta}_i)} .
\end{equation}
A microscopic calculation from the general action for our model, given in the next section, yields the following values for the Ginzburg-Landau coefficients:
\begin{equation}
\begin{array}{cccc}
 \alpha_1 =  \frac{p_{FB}^2}{48 \pi T_c^2} \frac{T-T_c}{T_c} ,  & \alpha_2 = \frac{1}{32 \pi T_c^2} , \\
 \beta = \frac{p_{FB}^2 }{1920 \pi T_c^4} ,  & \gamma  = \frac{\gamma_1^2 p_{FB}^2}{8 \pi T_c^4 N^5} .
\end{array}
\label{glcoeffs}
\end{equation}
From these coefficients one can extract a coherence length $\xi^2 \sim \alpha_2 / |\alpha_1| = 3 T_c / (2 p_{FB}^2  \vert T_c -T \vert)$, which is very short ($\xi \sim v_F/\gamma_1 \ll v_F/\Delta$) for $T$ sufficiently below $T_c$ indicating the strong coupling nature of the superconductivity.

To consider fluctuations for $T<T_c$, we expand around the mean-field value of $\Delta_{0}^2 = -(\alpha_1 - \gamma) / (2 \beta)$ as $\Delta_i(x) = (\Delta_0 + \delta_i(x)) e^{i \phi_i(x)}$. The fluctuation part of the action can then be written in terms of four fluctuation modes:
\begin{eqnarray}
S_1 & = & \int d^2x \frac{\alpha_2}{4 \beta} \vert \alpha_1-\gamma \vert (\nabla \phi_+)^2 \nonumber \\
S_2 & = & \int d^2x \left[\frac{\alpha_2}{2} (\nabla \delta_+)^2 + \vert \alpha_1 - \gamma \vert \delta_+^2\right] \nonumber \\
S_3 & = & \int d^2x \left[ \frac{\gamma \vert \alpha_1 - \gamma \vert}{2 \beta} \phi_-^2 + \frac{\alpha_2 \vert \alpha_1 - \gamma \vert}{4 \beta} (\nabla \phi_-)^2 \right] \nonumber \\
S_4 & = & \int d^2 x \left[\frac{\alpha_2}{2} (\nabla \delta_-)^2 + \vert \alpha_1 - 2 \gamma \vert \delta_-^2\right] 
\end{eqnarray}
so that $S_{GL} = S_{MF} + S_1 + S_2 + S_3 + S_4$, where the mean-field part is given by $S_{MF} = - (\alpha_1-\gamma)^2 /(2 \beta)$. The four fluctuation modes can be identified as the total phase (Nambu-Goldstone) mode ($S_1$), the total amplitude (Higgs-Anderson) mode ($S_2$), the relative phase (Leggett) mode ($S_3$), and the relative amplitude mode ($S_4$). In this new diagonal basis, the four independent fields are defined as $\phi_\pm = \phi_1 \pm \phi_N$ and $\delta_\pm = \delta_1 \pm \delta_N$. In the limit $N\rightarrow \infty$, the two phase and the two amplitude modes become identical and the action separates into two identical standard fluctuating Ginzburg-Landau actions with coefficients given by \eqref{glcoeffs}. The standard Ginzburg-Landau theory calculations can then be applied \cite{larkin2005theory} and we find that the total amplitude mode contribution to the heat capacity below $T_c$ becomes \cite{comment}
\begin{equation}
\delta C_{\rm{amplitude}} = \frac{A p_{FB}^2 T_c}{3 \pi (T_c - T)} \sim \frac{|\alpha_1|}{\alpha_2} \frac{T_c^2}{(T_c-T)^2} A.
\end{equation}
Here $A$ is the surface area of the sample. In the lowest order in $T_c - T$, the phase mode does not contribute to the heat capacity. Setting $\delta C_{\rm{amplitude}}(T^*)$ equal to the heat capacity jump at the transition,
\begin{equation}
\Delta C = A p_{FB}^2 \frac{5}{6 \pi} \sim \frac{\alpha_1^2}{\beta} \frac{T_c^2}{(T-T_c)^2} A,
\end{equation}
and solving for temperature, $T^*$, yields the Ginzburg number
\begin{equation}
Gi \equiv \frac{T_c - T^*}{T_c} = \frac{2}{5} .
\end{equation}
This is of the order of unity as expected from purely dimensional arguments. While at the reduced temperature indicated by the Ginzburg number, the Ginzburg-Landau approach is not, strictly speaking, valid, this nevertheless gives us an estimate of the size of the fluctuations. 

To understand the large fluctuations and compare them to conventional superconductors, we note that in both cases the fluctuation contribution and the heat capacity jump are proportional to the same Ginzburg-Landau coefficients, but the microscopic values of the coefficients differ greatly. In a conventional two-dimensional BCS superconductor \cite{larkin2005theory}
 $\alpha_1 = N(0) (T-T_c) / T_c^2$, 
 $\alpha_2 = 7 \zeta(3) N(0) v_F^2 / (32 \pi^2 T_C^3)$ and
 $\beta =7 \zeta(3) N(0) / (16 \pi^2 T_c^3)$,
where $\zeta(x)$ is the Riemann zeta function, $N(0)$ is the density of states at the Fermi level and $v_F$ is the Fermi velocity. There is one additional parameter compared to the case of the flat band model. This leads to different scales for the heat capacity jump and the fluctuation heat capacity in two-dimensional BCS superconductors
\begin{equation}
\delta C_{\rm{amplitude}} \sim \frac{T_c}{E_F} \frac{T_c}{|T-T_c|} \Delta C \label{BCSCamp}
\end{equation}
resulting in a much smaller Ginzburg number $Gi \sim  T_c / E_F$ compared to the flat band superconductors. 

\section{Microscopic calculation and arbitrary temperatures}

At low temperatures, the Ginzburg-Landau approach is invalid and we need a microscopic theory to account for the fluctuations. For the surface state, we can use the Bogoliubov - de Gennes equation derived in Ref.~\onlinecite{PhysRevB.83.220503} to deduce that the action has the form
\begin{eqnarray}
S &= & \sum_{p,p'}
\check{\bar{\Psi}}(p)
\begin{pmatrix}
\check{\Delta}_{1,p-p'} - i\tilde{\omega}_p & \check{\tau}_3 \xi_{\mathbf{p}}  \\
 \check{\tau}_3 \xi_{\mathbf{p}} & \check{\Delta}_{N,p-p'} - i\tilde{\omega}_p
\end{pmatrix}
\check{\Psi}(p') \nonumber \\
&& + \frac{V}{g T}\sum_p (\vert \Delta_{1,p} \vert^2 + \vert \Delta_{N,p} \vert^2 ) \nonumber \\
&\equiv & \sum_{p,p'} \check{\bar{\Psi}}(p) \mathcal{G}^{-1}_{p,p'} \check{\Psi}(p') + \frac{V}{g T}\sum_p (\vert \Delta_{1,p} \vert^2 + \vert \Delta_{N,p} \vert^2 ). \nonumber \\
\label{action1}
\end{eqnarray}
Here $\xi_{\mathbf{p}} = \gamma_1 (\vert \mathbf{p}\vert / p_{FB})^N$, $\tilde{\omega}_p = (2n+1) \pi T /(1-\mathbf{p}^2 /p_{FB}^2)$ is the Matsubara frequency with a momentum dependent factor which comes from using the ansatz for surface state wave functions given in Ref.~\onlinecite{PhysRevB.83.220503}, and $g$ is the superconducting coupling strength with dimensions of $(\rm{energy})\cdot(\rm{volume})$. The volume $V \equiv A d$, where $d \sim 1/p_{\rm FB}$ is the spatial extent of the surface state wavefunctions. Here, $p\equiv (n, \mathbf{p})$ is the three-momentum which includes the two-dimensional momentum $\mathbf{p}$ in the plane of the layers and the Matsubara index $n$. The $4\times 4$ matrix structure of the action (in addition to the momentum space degrees of freedom) results from top-bottom layer $\otimes$ particle-hole degrees of freedom. The Cooper pairs are described by matrices
\begin{equation}
\hat{\Delta}_{i,p-p'} = \begin{pmatrix}
0 & \Delta_{i,p-p'} \\
\bar{\Delta}_{i,p'-p} & 0 
\end{pmatrix}
\end{equation}
and can now have a non-zero momentum, i.e., $p-p' \neq 0$.

Integrating over the fermionic modes in \eqref{action1} results in
\begin{eqnarray}
\mathcal{Z} = \int \mathcal{D} \Delta \mathcal{D} \bar{\Delta} \exp\Bigg[ && - \frac{V}{g T}\sum_p (\vert \Delta_{1,p} \vert^2 + \vert \Delta_{N,p} \vert^2 ) \nonumber \\
&& + \rm{tr}\ln\left(\mathcal{G}^{-1}/T\right) \Bigg],
\label{partition}
\end{eqnarray}
where the trace $\rm{tr}$ is taken over the three-momentum and matrix indices.

From expression \eqref{partition}, the Ginzburg-Landau action, \eqref{glaction}, with the coefficients \eqref{glcoeffs} can be obtained by expanding in small $\Delta_{i,p}$. The mean-field value for the Cooper pair field $\Delta_0(T)$ is found by minimizing the action and $T_c$ is found by solving for the highest temperature below which the mean-field value is non-zero. In the limit $N \to \infty$, this yields the mean-field value $\Delta_0(0)= g p_{FB}^2 / (16 d \pi)$ and $T_c = \Delta_0(0)/ 3$ which we have used in expressing the Ginzburg-Landau coefficients in terms of the critical temperature. More generally, we can expand $\Delta_i$ around this mean-field value, $\Delta_i = \Delta_0 + \delta \Delta_i$. This separates the action into a mean-field part and a fluctuation part, $S = S_{MF}+S_\delta$, where the fluctuation part is given by
\begin{widetext}
\begin{equation}
S_\delta = \frac{1}{2} \sum_q
\overrightarrow{\delta \Delta}_q^\dagger
\begin{pmatrix}
-\mathcal{W}_q + \frac{V}{g T} & \mathcal{D}_q & - \mathcal{X}_q & 0 \\
\mathcal{D}_q & -\mathcal{W}_{-q} + \frac{V}{g T} & 0 & -\mathcal{X}_{-q} \\
-\mathcal{X}_q & 0 & -\mathcal{W}_q + \frac{V}{g T} & \mathcal{D}_q \\
0 & -\mathcal{X}_{-q} & \mathcal{D}_q & -\mathcal{W}_{-q} + \frac{V}{g T}
\end{pmatrix}
\overrightarrow{\delta \Delta}_q 
\label{fluctaction}
\end{equation}
\end{widetext}
and the mean-field part by
\begin{equation}
S_{MF} = -4 \sum_{\mathbf{p}} \ln\left[ \cosh\big(E_\mathbf{p}/2T\big)\right] +\frac{2 V}{g T} \vert \Delta_0 \vert^2.
\label{eqSMF}
\end{equation}
In the latter we have performed the summation over the Matsubara frequencies and $E_\mathbf{p}$ are the positive quasiparticle energies given by Eq.~(\ref{spectrum1}).
The partition function is then given as an integral over the fluctuation modes as
\begin{equation}
\mathcal{Z} = e^{-S_{MF}} \int \mathcal{D} \overrightarrow{\delta \Delta} \mathcal{D} \overrightarrow{\delta \Delta}^\dagger e^{-S_\delta} .
\end{equation}
We have here written the action in a convenient matrix form with the vector fluctuation fields given by $\overrightarrow{\delta \Delta}_q = (\delta \Delta_{1,q}, \bar{\delta \Delta}_{1,-q}, \delta \Delta_{N,q}, \bar{\delta \Delta}_{N,-q})^T$. The matrix elements in the fluctuation action are given in terms of three polarization operators. Expressions for them are
\begin{eqnarray}
\mathcal{W}_q & =& \sum_p \frac{\tilde{\omega}_p \tilde{\omega}_{p-q}}{(\tilde{\omega}_p^2 + \Delta_0^2 + \xi_p^2)(\tilde{\omega}_{p-q}^2 + \Delta_0^2 + \xi_{p-q}^2)} \nonumber \\
\mathcal{D}_q & = &\sum_p \frac{\Delta_0^2}{(\tilde{\omega}_p^2 + \Delta_0^2 + \xi_p^2)(\tilde{\omega}_{p-q}^2 + \Delta_0^2 + \xi_{p-q}^2)} \nonumber \\
\mathcal{X}_q & =& \sum_p \frac{\xi_p \xi_{p-q}}{(\tilde{\omega}_p^2 + \Delta_0^2 + \xi_p^2)(\tilde{\omega}_{p-q}^2 + \Delta_0^2 + \xi_{p-q}^2)} .
\end{eqnarray}
As for the Ginzburg-Landau action, the four eigenmodes of the fluctuation action correspond to relative and total phase fluctuations and relative and total amplitude fluctuations. Also, again, in the limit of a large number of layers, only the total amplitude and total phase modes survive and the third polarization operator, $\mathcal{X}_q$, vanishes.

\begin{figure}[!ht]
\centering
\includegraphics[width=.99\columnwidth]{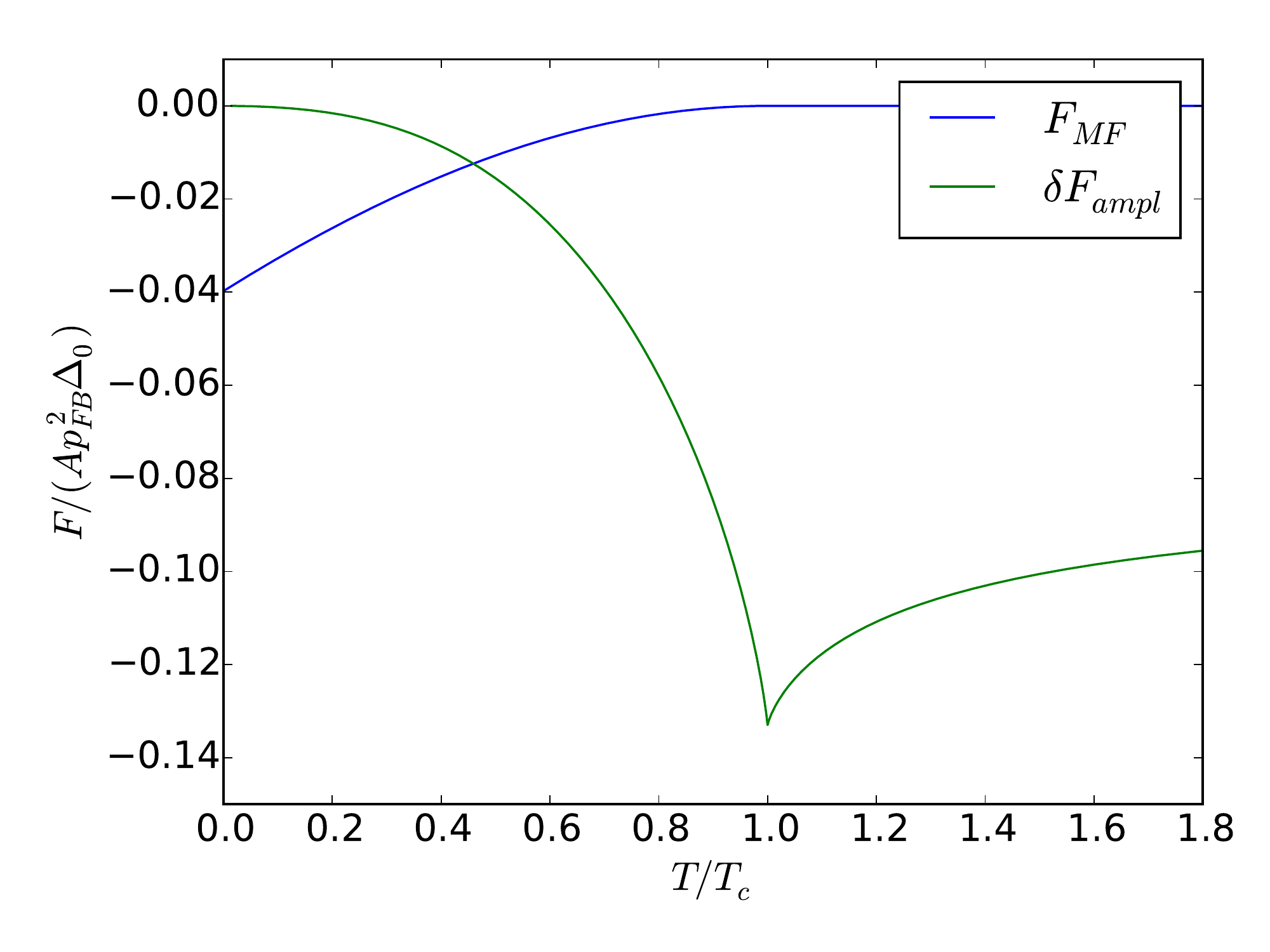}
\caption{\small Free energy contribution from the amplitude fluctuations and from the mean-field calculation.}
\label{fig:dF}
\end{figure}

From the Ginzburg-Landau calculation, we know that the dominating contribution to the thermodynamics in the large $N$ limit comes from the amplitude fluctuation part. Taking into account only this mode, the fluctuation contribution to the free energy, $\delta F = -T \ln \delta \mathcal{Z} \approx \delta F_{\rm{amplitude}}$, becomes
\begin{equation}
\delta F_{\rm{amplitude}} = 2 T \sum_q \ln\left(1- \frac{\mathcal{W}_q - \mathcal{D}_q}{V / g T} \right) .
\end{equation}
We calculate $\delta F_{\rm{amplitude}}$ as well as the heat capacity, $\delta C_{\rm{amplitude}} = -T \partial_T^2 \delta F_{\rm{amplitude}}$, numerically. The result for the free energy is shown in Fig. \ref{fig:dF}. For zero temperature, the fluctuation contribution goes to zero, but grows larger than the mean-field contribution as the temperature increases. Above the critical temperature, $\delta F_{\rm{amplitude}}$ saturates into a finite value. 

\begin{figure}[!ht]
\centering
\includegraphics[width=.99\columnwidth]{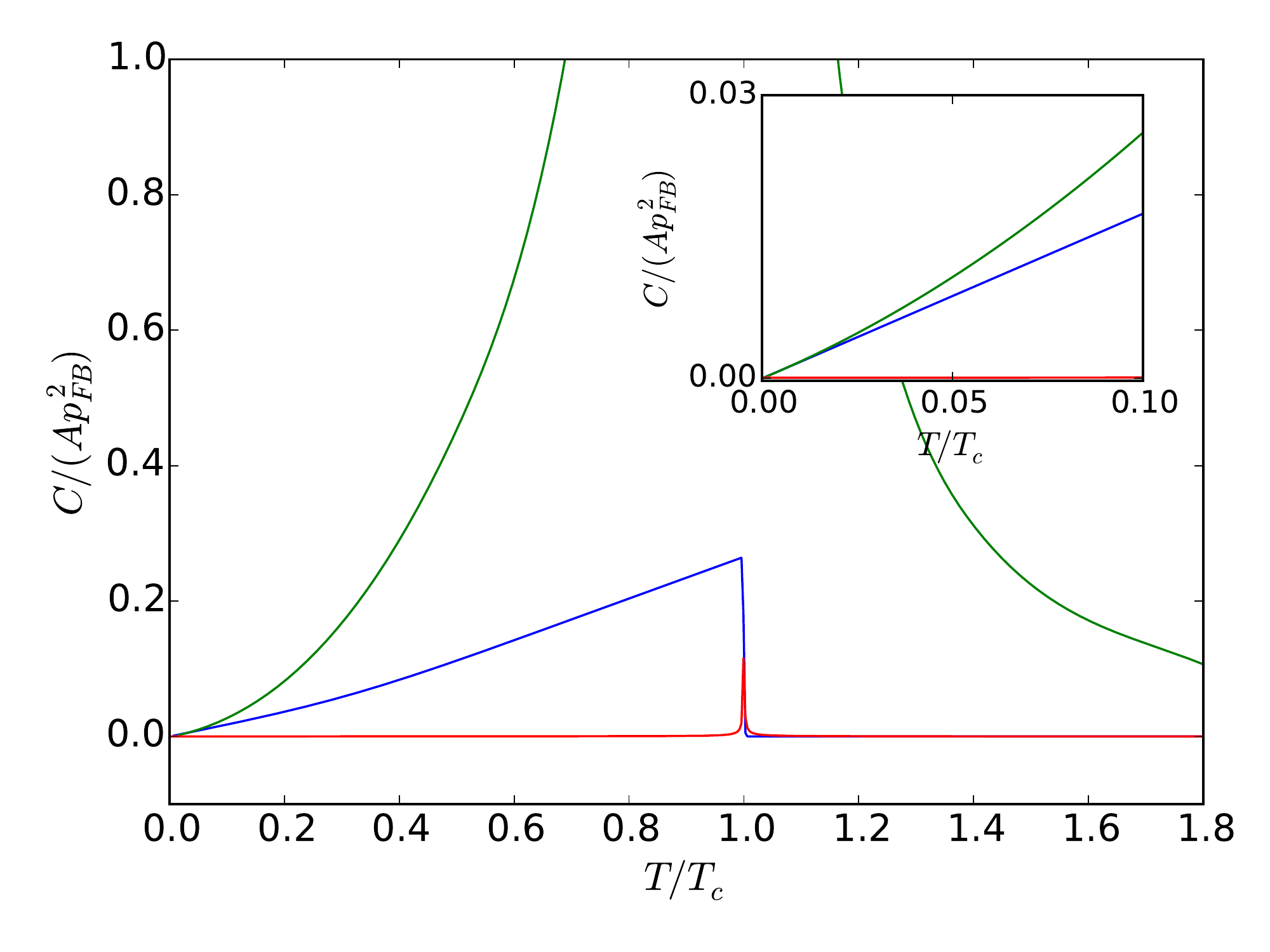}
\caption{\small Specific heat contribution from the amplitude fluctuations (green curve) and from the mean-field calculation (blue curve). For comparison the fluctuation contribution for a conventional BCS superconductor with $T_c = 10^{-4} E_F$ and equal heat capacity jump at the transition $\Delta C$  is also shown (red curve). Inset shows the low-temperature behaviour of the three curves.}
\label{fig:dC}
\end{figure}

For the heat capacity, the numerical result is shown in Fig. \ref{fig:dC}. At low temperatures we  find that the fluctuation contribution to the heat capacity becomes equal to the mean field heat capacity $C_{MF}$, both given by
\begin{equation}
C(T\ll T_c) = \frac{A p_{FB}^2 \pi}{6} \frac{T}{\Delta_0}. \label{lowTC}
\end{equation}
For  $C_{MF}$ Eq.~(\ref{lowTC}) straightforwardly follows from Eq.~(\ref{eqSMF}). On the other hand, the fact that $\delta C_{\rm{amplitude}}$ is also given by Eq.~(\ref{lowTC}) can be analytically understood by assuming that  it  comes from a free boson contribution corresponding to the amplitude mode dispersion, which can be determined from the corresponding eigenvalue of the matrix in the fluctuation action \eqref{fluctaction}. Namely, this way we find that 
for $T=0$ and $N \to \infty$ the amplitude mode  dispersion is given by
\begin{equation}
E_{\mathbf{p}, \rm{ampl}}(T=0) = \Delta_0 \mathbf{p}^2/p_{FB}^2,
\label{amplspec}
\end{equation}
and the corresponding heat capacity is indeed given by Eq.~(\ref{lowTC}).
For higher temperatures, the fluctuation contribution completely dominates over the mean field contribution. At $T_c$, the heat capacity diverges as $\sim 1/\vert T - T_c \vert$ as predicted by the Ginzburg-Landau theory. The temperature at which the fluctuation heat capacity equals the mean-field heat capacity jump at the transition is approximately $T/T_c \approx 0.4$ which is on the same order as expected from the Ginzburg number. For comparison, we also show the Ginzburg-Landau fluctuation contribution expected for a conventional BCS superconductor  [Eq.~(\ref{BCSCamp})] with $T_c / E_F \sim 10^{-4}$ and other parameters chosen so that the mean-field heat capacity jump at the transition $\Delta C$ would be the same as the one we get for the flat band case.

\section{Discussion}

The large contribution from the amplitude mode even at low temperatures can be understood from the form of the amplitude mode dispersion at $T=0$. By investigating the eigenvalues of the matrix in the fluctuation action, Eq. \eqref{fluctaction}, one can show that the mass gap of the mode is not $2 \Delta_0$ as in conventional superconductors, but instead it is $2 E_{\mathbf{p},min}$, where $E_{\mathbf{p},min}$ is the minimum of the fermionic spectrum. Since the spectrum of the fermionic excitations is given by Eq. \eqref{spectrum1}, which does not have a gap in the limit of large $N$ (see Fig. \ref{fig:spectrum}), also the gap in the amplitude mode closes and the contribution from the fermionic mode and the bosonic amplitude mode can be of the same order of magnitude at low temperatures. Therefore, finite $N$ corrections to our result would likely diminish the fluctuation contribution, improving the validity range of the mean-field theory.

Besides rhombohedral graphite, there have been also other suggestions for correlated states at the surfaces of topological semimetals. We expect that their mean-field spectrum is also gapless in the bulk limit, and probably exhibits very similar fluctuation contributions as that described in this paper. Moreover, flat bands can also emerge at the surface in superconductors with a nodal order parameter structure possibly resulting in correlated states\cite{schnyder2012types, honerkamp2000instabilities, covington1997observation, potter2014edge, li2013spontaneous}. Similarly as in the present situation, the fluctuation contribution to the heat capacity may also dominate the mean field contribution in these correlated states, so that the mean field theory becomes inaccurate and the interactions lead to a strongly correlated state.

We consider here only the effect of the amplitude mode to thermodynamics since we assume that it dominates over the phase mode contribution. At reasonably high temperatures it follows from general Ginzburg-Landau theory that the amplitude mode contribution dominates because of its singularity at $T=T_c$. At lower temperatures,  the phase mode contribution to thermodynamic properties in a charged system is also influenced by the coupling to the electromagnetic field \cite{anderson1958random, Anderson63}. In the calculation of this coupling it should be taken into account that although the order parameter appears only close to the surface the supercurrent flows also in the bulk \cite{Kopnin2011surface}. 

Lastly, we want to emphasize that we do not consider the coupling of the fluctuations to external fields. The large contribution from the amplitude fluctuations to the heat capacity might or might not appear in other observables depending on whether they couple to this degree of freedom or not. For example, it is known that the electromagnetic field does not couple to the amplitude mode at $T=0$ directly \cite{littlewood1982amplitude} and thus such a large fluctuation effect is not necessarily present in the current.

\section{Acknowledgements}

We thank G.E. Volovik for helpful discussions. This work was supported by the Academy of Finland through its Center of Excellence program, and by the European Research Council (Grant No. 240362-Heattronics).


\begin{thebibliography}{31}
\bibitem [{cut()}]{cutoff}%
  \BibitemOpen
  \href@noop {} {}\bibinfo {note} {We use $|\mathbf{p}| = (1- 1/N) p_{FB}$ as a
  cut-off momentum in all integrals.}\BibitemShut {Stop}%
\bibitem [{foo()}]{footnote1}%
  \BibitemOpen
  \href@noop {} {}\bibinfo {note} {We use term Cooper pair field for the
  fluctuating complex fields $\Delta_{1,N}$ in the action. The corresponding
  mean field is the pair potential.}\BibitemShut {Stop}%
\end{thebibliography}
\end{document}